\documentclass[a4paper]{jpconf}
\usepackage{graphicx}
\usepackage{xspace}

\def\modell{{\sl poly-gonato} model\xspace}
\def\knee{{\sl knee}\xspace}
\def\knees{{\sl knees}\xspace}
\def\Xmax{$X_{max}$\xspace}
\def\gcm2{g/cm$^2$}
\def\lnA{\langle\ln A\rangle}
\def\Cerenkov{\v{C}erenkov\xspace}

\def\fref#1{Fig.~\ref{#1}}

\def\sref#1{Sect.~\ref{#1}}

\def\xx{\vspace*{-3mm}}
\def\yy{0.933}

\begin{document}
\title{A review of experimental results at the knee%
       \footnote{Invited talk, presented at the "Workshop on Physics of the End
       of the Galactic Cosmic Ray Spectrum", Aspen, USA, April 25 - 29, 2005}}

\author{J\"org R. H\"orandel}

\address{Institute for Experimental Nuclear Physics, University of Karlsruhe,
         P.O. Box 3640, 76021 Karlsruhe, Germany}

\ead{hoerandel@ik.fzk.de}

\begin{abstract}
 Results of experiments investigating air showers in the energy region of the
 knee are summarized. The all-particle energy spectrum, the mean logarithmic
 mass, and the average depth of the shower maximum will be discussed. 
 Spectra for groups of elements from air shower data are compared to 
 results from direct measurements.
\end{abstract}

\section{Introduction}
One of the most remarkable structures in the energy spectrum of cosmic rays is
a change of the spectral index $\gamma$ of the power law $dN/dE\propto
E^\gamma$ at an energy of about 4~PeV, the so called \knee.  In the literature
various reasons for the origin of the \knee are discussed \cite{origin}. Some
popular concepts are: The maximum energy attained during the acceleration at
supernova remnant shock waves
\cite{berezhko,stanev,kobayakawa,sveshnikova,wolfendale}, reacceleration in the
galactic wind \cite{voelk}, acceleration in pulsars \cite{bednarek}, or leakage
of particles from the Galaxy during the propagation process
\cite{ptuskin,kalmykov,ogio,roulet,swordy,lagutin}. Different scenarios are the
acceleration of cosmic rays in $\gamma$-ray bursts \cite{plaga,wick,dar}.
During the propagation through the Galaxy cosmic-ray particles are proposed to
interact with dense photon fields close to the sources or with background
neutrinos \cite{tkaczyk,dova,wigmans,candia}.  As a last possibility, new
processes between elementary particles in the atmosphere are postulated which
may transfer energy into non-observed channels and thus cause a \knee in the
observed air shower components \cite{nikolsky,petrukhin,kazanas}.

An answer to the question of the cause for the \knee is expected to reveal also
crucial information on the origin of galactic cosmic rays. Experimental access
to such questions is provided by measurements of charged cosmic rays (the
classical nucleonic component) and $\gamma$-rays with experiments above the
atmosphere and by the observation of air showers initiated by high-energy
particles in the atmosphere.

A wealth of information on potential cosmic-ray sources is provided by recent
measurements of TeV $\gamma$-rays from shell type supernova remnants with the
H.E.S.S. experiment \cite{hesssnr}. The observations reveal a shell structure of
the remnant and an energy spectrum of $\gamma$-rays $\propto E^{-2.2}$ in
agreement with the idea of particle acceleration in the shock front.  The
spectrum extends up to energies of 10~TeV and provides evidence for the
existence of particles with energies beyond 100~TeV at the shock front that
emerged from the supernova explosion.

The gyromagnetic radius of a proton with an energy of 1~PeV in the galactic
magnetic fields is about 0.4~pc and consequently the probability to detect a
source in charged galactic cosmic rays is rather small. Indeed, observations by
the KASCADE experiment did not reveal any significant evidence for point
sources in the energy range from 0.3 to about 100~PeV \cite{kascade-points}.
Nevertheless, information on the sources is derived from detailed measurements
of the cosmic-ray composition.  Investigations of the abundance of refractory
nuclides reveal that their abundance at the sources is extremely similar to the
abundance observed in the solar system \cite{cris-abundance}. This indicates
that cosmic rays are accelerated from a well-mixed sample of contemporary
interstellar matter.

Information on the propagation of particles through the Galaxy is obtained by
measurements of the ratio from primary to secondary nuclei, the latter are
produced in spallation reactions of cosmic rays with particles of the
interstellar medium, and the abundance of radioactive nuclides in cosmic rays.
Measurements by the CRIS experiment yield a residence time for cosmic rays in
the Galaxy of $15\cdot10^6$~yrs and a propagation path length of about 10~\gcm2
for particles with GeV energies \cite{cris-time}.  The measured abundances at
GeV energies are frequently described using leaky-box models.  At higher
energies measurements of anisotropy amplitudes yield information on the
propagation process \cite{kascade-aniso}. They indicate that the propagation 
can be described by diffusion models, while leaky-box models exhibit too
large anisotropy values even at relatively low energies.

At PeV and EeV energies, information is derived from the measurements of the
energy spectrum and mass composition of cosmic rays.
In the following, experimental results are reviewed on the 
all-particle energy spectrum (\sref{espek}),
the average mass composition (\sref{comp}),
and spectra for elemental groups (\sref{elementspek}).

\section{All-particle energy spectrum} \label{espek}

\begin{figure}[t] \centering
  \includegraphics[width=\yy\textwidth]{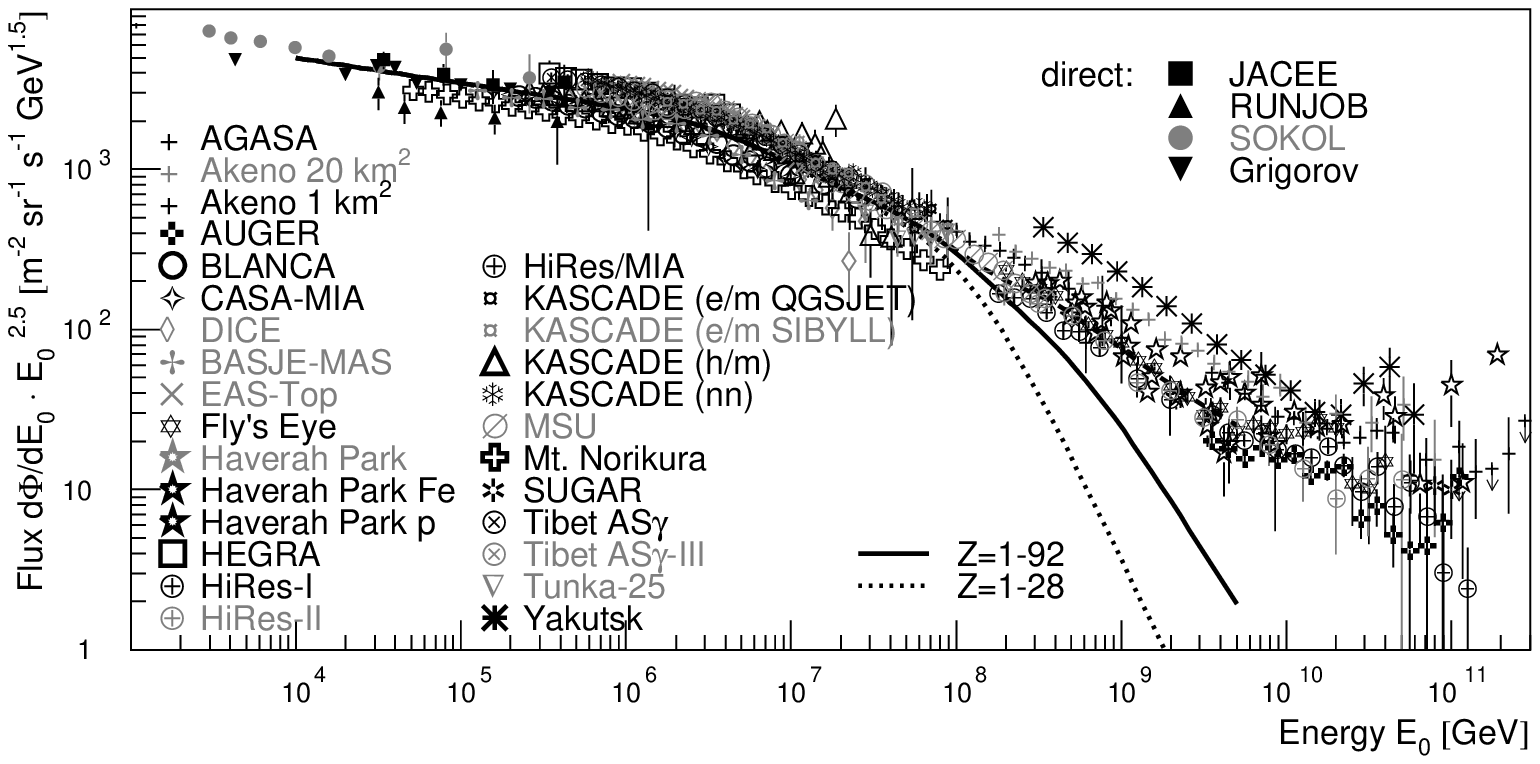} \xx 
  \caption{All-particle energy spectrum. Results from direct measurements by
   Grigorov \etal \cite{grigorov},
   JACEE \cite{jaceefe},
   RUNJOB \cite{runjob05}, and
   SOKOL \cite{sokol}
   as well as from the air shower experiments
   AGASA \cite{agasa}
   Akeno 1~km$^2$ \cite{akeno1}
   and  20~km$^2$ \cite{akeno20},
   AUGER \cite{auger05},
   BASJE-MAS \cite{basjemas},
   BLANCA \cite{blanca},
   CASA-MIA \cite{casae},
   DICE \cite{dice},
   EAS-TOP \cite{eastope},
   Fly's Eye \cite{flyseye},
   Haverah Park (1991) \cite{haverahpark91}
   and (2003) \cite{haverahpark03},
   HEGRA \cite{hegraairobic},
   HiRes-MIA \cite{hiresmia},
   HiRes-I \cite{hiresi},
   HiRes-II \cite{hiresii},
   KASCADE electrons and muons interpreted with two hadronic interaction models
        \cite{ulrichapp}, 
	hadrons \cite{hknie}, and a neural network analysis combining
	different shower components \cite{rothnn},
   MSU \cite{msu},	
   Mt.~Norikura \cite{mtnorikura},
   SUGAR \cite{sugar},
   Tibet AS$\gamma$ \cite{tibetasg00} and 
         AS$\gamma$-III \cite{tibetasg03},
   Tunka-25 \cite{tunka04}, and
   Yakutsk \cite{yakutsk5001000}.
   The lines indicate the spectrum according to the \modell.}
   \label{allpart}
\end{figure}

The all-particle energy spectra obtained by many experiments are compiled in
\fref{allpart}. Shown are results from direct measurements above the atmosphere
as well as from various air shower experiments. The individual measurements
agree within a factor of two in the flux values and a similar shape can be
recognized for all experiments with a \knee at energies of about 4~PeV.
Typical values for the systematic uncertainties of the absolute energy scale
for air shower experiments are about 15 to 20\%.  Renormalizing the energy
scales of the individual experiments to match the all-particle spectrum obtained
by direct measurements in the energy region up to almost a PeV requires
correction factors in the order of $\pm10$\% \cite{pg}.  A remarkable result,
indicating that behind an absorber of 11 hadronic interaction
lengths or 30 radiation lengths the energy of the primary particle is 
determined with an absolute error in the order of $\pm10$\%.  One should keep
in mind that the experiments investigate different air shower components,  are
situated at different atmospheric depths, and use different interactions models
to interpret the observed data. Nevertheless, the systematic differences are
relatively small and the all-particle spectrum seems to be well known. Average
experimental fluxes are tabulated in \cite{pg}.

The lines in \fref{allpart} indicate sum spectra obtained by extrapolations of
the energy spectra for individual elements from direct measurements assuming
power laws with a cut-off proportional to the charge of the respective element
according to the \modell \cite{pg}.  Sum spectra for elements from hydrogen to
nickel and for all elements are shown. The extrapolated all-particle flux is
compatible with the flux derived from air shower experiments in the \knee
region. Above $10^8$~GeV the flux of galactic cosmic rays is not sufficient to
account for the observed all particle spectrum, and an additional, presumably
extragalactic component is required.

\section{Average mass of cosmic rays} \label{comp}

\begin{figure}[t] \centering
  \includegraphics[width=\yy\textwidth]{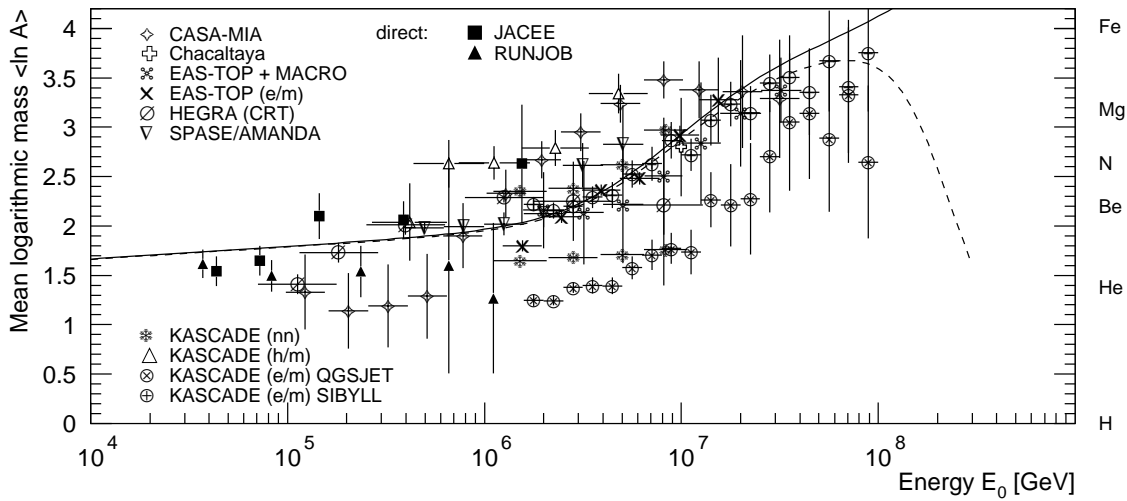} \xx
  \caption{Mean logarithmic mass of cosmic rays derived from the measurements
           of electrons, muons, and hadrons at ground level. Results are shown
	   from
	   CASA-MIA \cite{casam},
	   Chacaltaya \cite{chacaltaya},
	   EAS-TOP electrons and GeV muons \cite{eastop-knee},
	   EAS-TOP/MACRO (TeV muons) \cite{eastop-macro-lna},
	   HEGRA CRT \cite{hegracrt},
	   KASCADE electrons and muons interpreted with two hadronic
	   interaction models \cite{ulrichapp},
	   hadrons and muons \cite{kascadehm}, 
	   as well as an analysis combining different observables with a 
	   neural network \cite{rothnn}, and
	   SPASE/AMANDA \cite{spaseamandalna}.
	   The lines indicate expectations according to the \modell.
	   For comparison, results from direct measurements are shown as well
	   from the JACEE \cite{jaceemasse} and RUNJOB \cite{runjob05}
	   experiments.
	   }
  \label{masse}	   
\end{figure}

At energies below a PeV energy spectra for individual elements have been
observed above the atmosphere \cite{wiebel}. At higher energies this is
presently not possible due to the low flux values and the large fluctuations in
the development of extensive air showers. Thus, mostly the mean mass is
investigated.  An often-used quantity to characterize the composition above
1~PeV is the mean logarithmic mass, defined as $\lnA= \sum_i r_i \ln A_i$,
$r_i$ being the relative fraction of nuclei of mass $A_i$.  In the
superposition model of air showers, the shower development of nuclei with mass
$A$ and energy $E_0$ is described by the sum of $A$ proton showers of energy
$E=E_0/A$.  The shower maximum $t$ penetrates into the atmosphere as $t
\propto\ln E$, hence, most air shower observables at ground level scale
proportional to $\ln A$. 

Frequently, the ratio of the number of electrons and muons is used to determine
the mass composition. The experiments CASA-MIA \cite{casam}, EAS-TOP
\cite{eastop-knee}, or KASCADE use muons with an energy of several 100~MeV to
1~GeV.  To study systematic effects two hadronic interaction models are used to
interpret the data measured with KASCADE \cite{ulrichapp}.  High energy muons
detected deep below rock or antarctic ice are utilized by the EAS-TOP/MACRO
\cite{eastop-macro-lna} and SPASE/AMANDA \cite{spaseamandalna} experiments.
Also the correlation between the hadronic and muonic shower components has been
investigated by KASCADE \cite{kascadehm}.  The production height of muons has
been reconstructed by HEGRA/CRT \cite{hegracrt} and KASCADE \cite{buettner}.

Results from various experiments measuring electrons, muons, and hadrons at
ground level are compiled in \fref{masse}.  At low energies the values for the
mean logarithmic mass are compared to results from direct measurements.  A
clear increase as function of energy can be recognized. The experimental values
follow a trend predicted by the \modell as indicated by the lines in the figure.
However, individual experiments exhibit systematic deviations of about $\pm1$
unit in $\lnA$ from the line.  Of particular interest are also the
investigations of the KASCADE experiment, interpreting the same measured data
with two different models for the interactions in the atmosphere results in a
systematic difference of about 0.7 to 1 in $\lnA$.

\begin{figure}[t] \centering
  \includegraphics[width=\yy\textwidth]{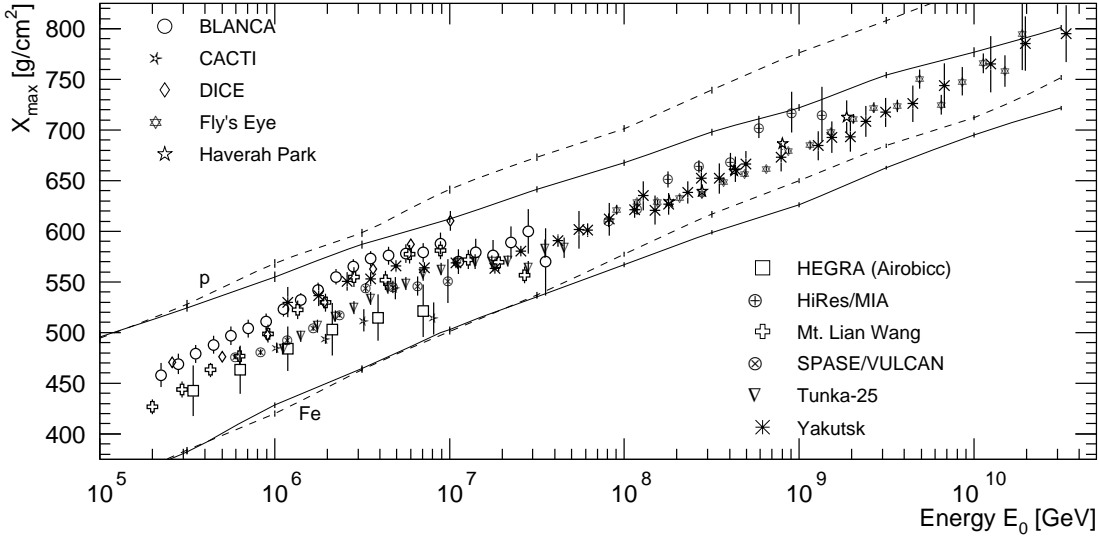} \xx
  \caption{Average depth of the shower maximum \Xmax as function of primary
	   energy as obtained by 
	   BLANCA \cite{blanca},
	   CACTI \cite{cacti},
	   DICE \cite{dice},
	   Fly's Eye \cite{flyseye},
	   Haverah Park \cite{haverahpark00},
	   HEGRA \cite{hegraairobic},
	   HiRes/MIA \cite{hires00},
	   Mt. Lian Wang \cite{mtlianwang},
	   SPASE/VULCAN \cite{spase99},
	   Tunka-25 \cite{tunka04}, and
	   Yakutsk \cite{yakutsk}. 
	   The lines indicate simulations for proton and iron induced showers
	   using the CORSIKA code with the hadronic interaction model QGSJET 01
	   (solid line) and a version with lower cross sections and slightly
	   increased elasticities (dashed line, model~3 in \cite{wq}).}
  \label{xmax} 
\end{figure}

Another technique to determine the mass of cosmic rays are measurements of the
average depth of the shower maximum using non-imaging and imaging \Cerenkov
detectors as well as fluorescence telescopes at the highest energies.  The
results of several experiments are presented in \fref{xmax} as function of
energy. The observed values are compared to predictions of air shower
simulations for primary protons and iron nuclei using the program CORSIKA
\cite{corsika} with the hadronic interaction model QGSJET~01 \cite{qgsjet} and
a modified version with lower cross sections and larger values for the
elasticity of the hadronic interactions (model~3a in \cite{wq}). The latter is
compatible with measurements at colliders, for details see \cite{wq}.  The
lower values for the total inelastic proton-antiproton cross sections yield
also lower inelastic proton-air cross sections, which are in good agreement
with recent measurements from the HiRes experiment
\cite{belovisvhecri,isvhecri04wq}.  In principle, the difference between the
solid and the dashed lines in the figure represents an estimate of the
projection of the experimental errors from collider experiments on the average
depth of the shower maximum in air showers.  At $10^9$~GeV the difference
between the two model versions for primary protons is about half the difference
between proton and iron induced showers. This illustrates the significance of
the uncertainties of the collider measurements for air shower observables.

\begin{figure}[t] \centering
  \includegraphics[width=\yy\textwidth]{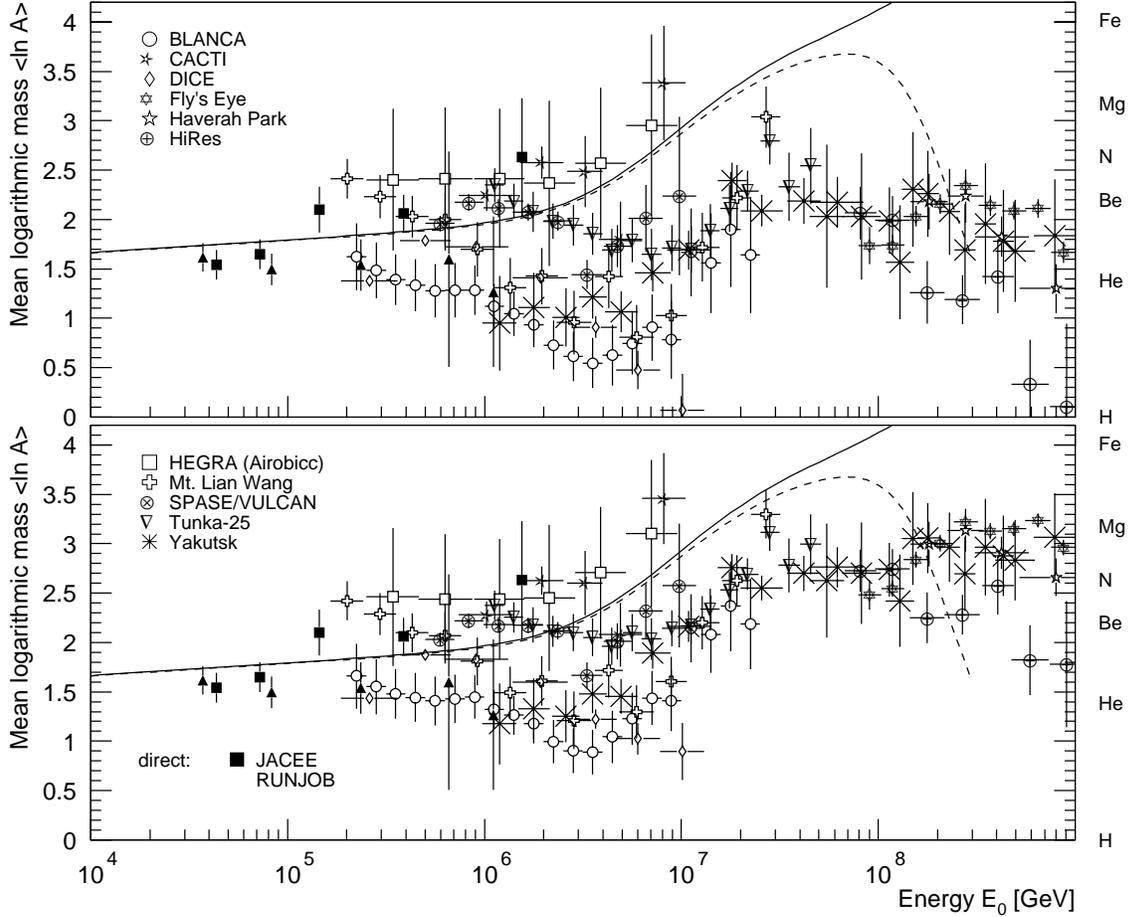} \xx
  \caption{Mean logarithmic mass of cosmic rays derived from the average
	   depth of the shower maximum, see \fref{xmax}.  Two hadronic
	   interaction models are used to interpret the measurements: QGSJET~01
	   (top) and a modified version with lower cross sections and a
	   slightly increased elasticity (model~3a \cite{wq}, bottom).  For
	   references, see caption of \fref{xmax}. For comparison, results
	   from direct measurements are shown as well from the JACEE
	   \cite{jaceemasse} and RUNJOB \cite{runjob05} experiments.
	   The lines indicate expectations according to the \modell.}
  \label{xmaxlna}	   
\end{figure}

Knowing the average depth of the shower maximum for protons $X_{max}^p$ and
iron nuclei $X_{max}^{Fe}$ from simulations, the mean logarithmic mass is
derived in the superposition model of air showers from the measured
$X_{max}^{meas}$ using
$\lnA=(X_{max}^{meas}-X_{max}^{p})/(X_{max}^{Fe}-X_{max}^{p}) \cdot \ln
A_{Fe}$.  The corresponding $\lnA$ values, obtained from the results shown in
\fref{xmax}, are plotted in \fref{xmaxlna} versus the primary energy using both
interaction models to interpret the observed data. Up to about a PeV there are
only marginal differences between the two interpretations.  On the other hand,
at large energies a significantly heavier composition is obtained for the
modified version (model~3a).  At $10^9$~GeV the differences amount to about 1
in $\lnA$.  These examples illustrate how strong the interpretation of air
shower measurements depends on model parameters such as the inelastic cross
sections or elasticities used. At Tevatron energies the cross sections vary
within the error range given by the experiments and at $10^8$~GeV the proton
air cross sections of QGSJET and model~3a differ only by about 10\%, but the
general trend of the emerging $\lnA$ distributions proves to be significantly
different.  
This underlines the importance to test and improve the understanding of
hadronic interactions in the atmosphere with air shower experiments
\cite{isvhecri02wwtest}.  It seems that the interpretation with model~3a is
better compatible with the mean logarithmic masses derived from electrons,
muons, and hadrons already shown in \fref{masse} and also with the predictions
of the \modell as indicated by the lines.

\section{Spectra for elemental groups} \label{elementspek}

\begin{figure}[t] \centering
  \includegraphics[width=\yy\textwidth]{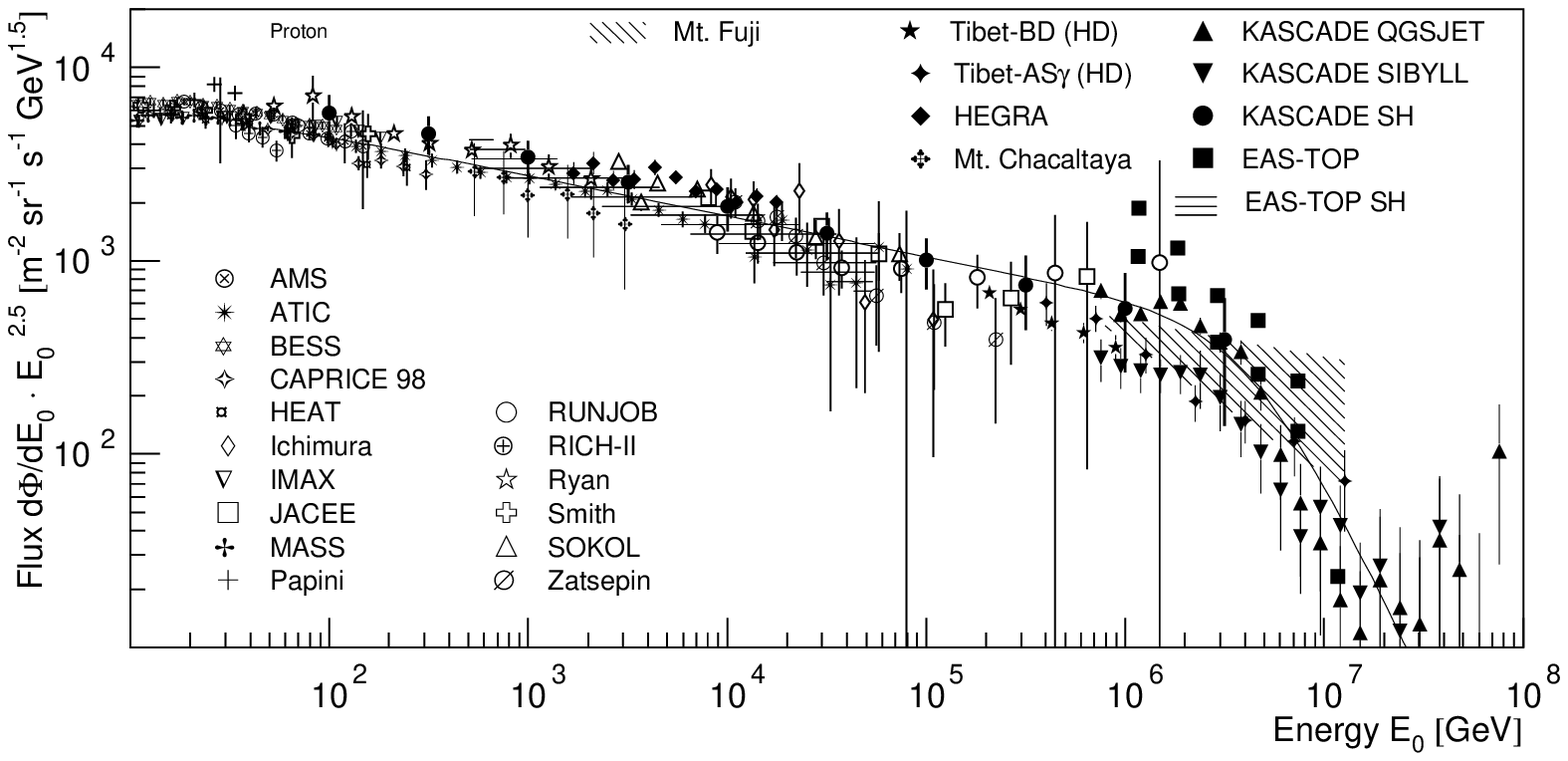} \xx
  \caption{Energy spectrum for protons.
   Results from direct measurements above the atmosphere by 
   AMS \cite{amsp},
   ATIC \cite{atic03},
   BESS \cite{bess00},
   CAPRICE \cite{caprice98},
   HEAT \cite{heat01},
   Ichimura \etal \cite{ichimura},
   IMAX \cite{imax00},
   JACEE \cite{jaceephe},
   MASS \cite{mass99},
   Papini \etal \cite{papini},
   RUNJOB \cite{runjob05},
   RICH-II \cite{rich2},
   Ryan \etal \cite{ryanp},
   Smith \etal \cite{smith},
   SOKOL \cite{sokol},
   Zatsepin \etal \cite{zatsepinp},
   and 
   fluxes obtained from indirect measurements by
   KASCADE electrons and muons for two hadronic interaction models
   \cite{ulrichapp} and single hadrons \cite{kascadesh},
   EAS-TOP (electrons and muons) \cite{eastopspec} and single hadrons
   \cite{eastopsh},
   HEGRA \cite{hegrap},
   Mt. Chacaltaya \cite{chacaltayap},
   Mts. Fuji and Kanbala \cite{mtfujip},
   Tibet burst detector (HD) \cite{tibetbdp} and AS$\gamma$ (HD) 
   \cite{tibetasgp}.
   The line indicates the spectrum according to the \modell.}
   \label{elementspekp}
\end{figure}

\begin{figure}[t] \centering
  \includegraphics[width=\yy\textwidth]{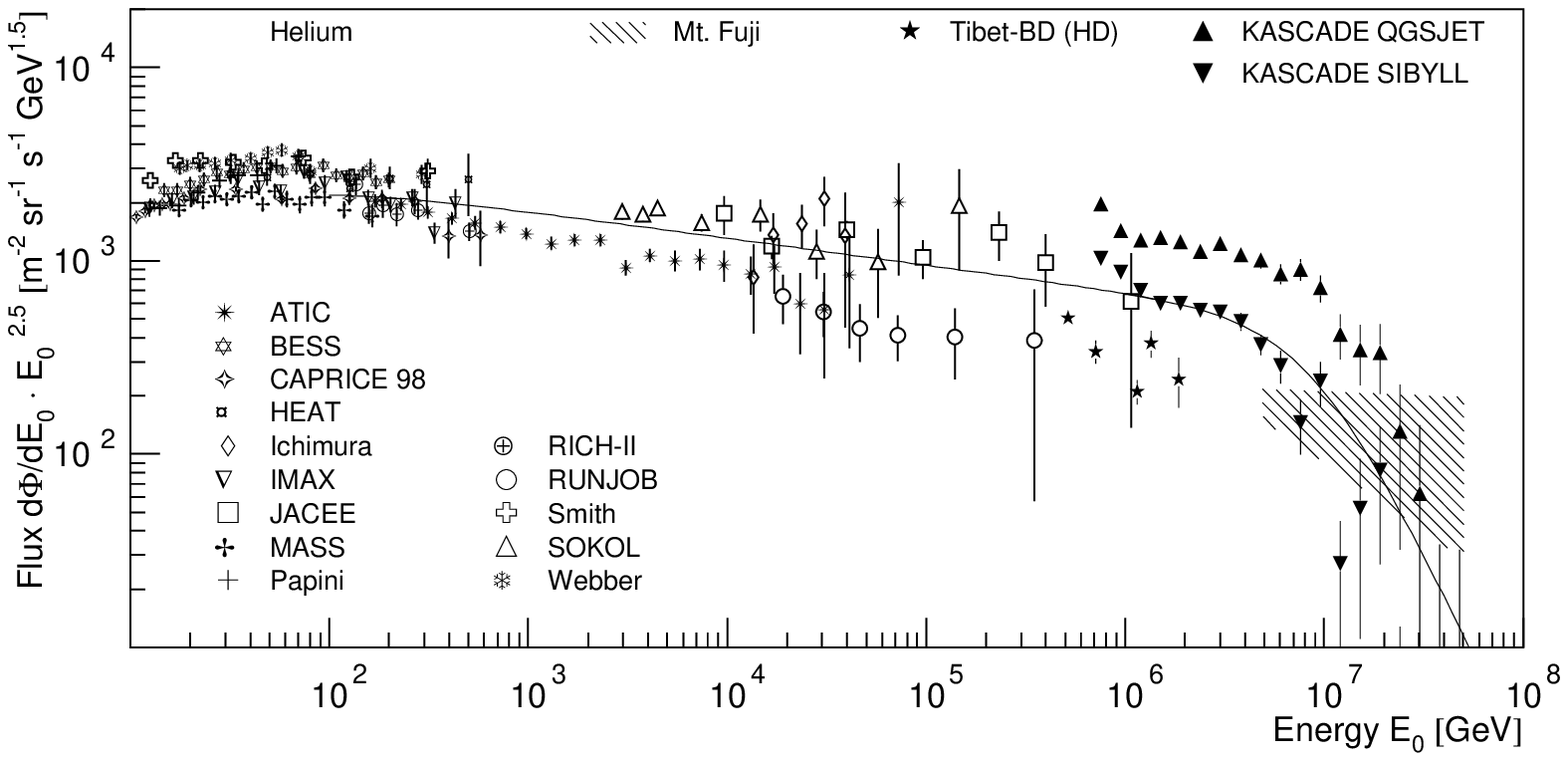} \xx
  \caption{Energy spectrum for helium nuclei.
   Results from direct measurements above the atmosphere by 
   ATIC \cite{atic03},
   BESS \cite{bess00},
   CAPRICE \cite{caprice98},
   HEAT \cite{heat01},
   Ichimura \etal \cite{ichimura},
   IMAX \cite{imax00},
   JACEE \cite{jaceephe},
   MASS \cite{mass99},
   Papini \etal \cite{papini},
   RICH-II \cite{rich2},
   RUNJOB \cite{runjob05},
   Smith \etal \cite{smith},
   SOKOL \cite{sokol},
   Webber \cite{webberhe},
   and 
   fluxes obtained from indirect measurements by
   KASCADE electrons and muons for two hadronic interaction models
   \cite{ulrichapp},
   Mts. Fuji and Kanbala \cite{mtfujip}, and
   Tibet burst detector (HD) \cite{tibetbdp}.
   The line indicates the spectrum according to the \modell.}
   \label{elementspekhe}
\end{figure}

\begin{figure}[t] \centering
  \includegraphics[width=\yy\textwidth]{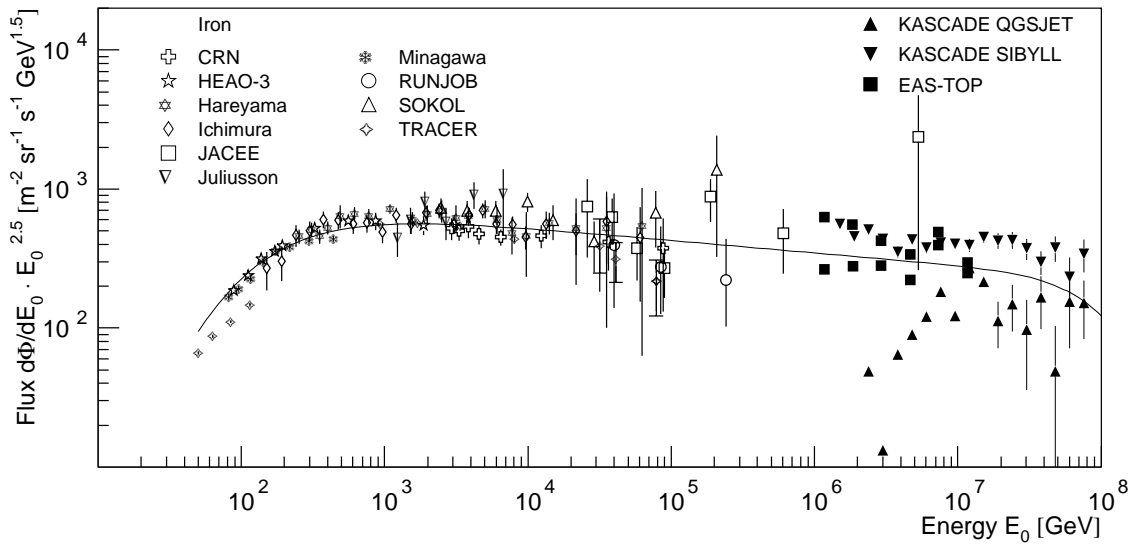} \xx
  \caption{Energy spectrum for iron nuclei.
   Results from direct measurements above the atmosphere by 
    CRN \cite{crn},
    HEAO-3 \cite{heao3},
    Juliusson \etal \cite{juliusson},
    Minagawa \etal \cite{minagawa},
    TRACER \cite{tracer05}
    (single element resolution) and
    Hareyama \etal \cite{hareyama},
    Ichimura \etal \cite{ichimura},
    JACEE \cite{jaceefe},
    RUNJOB \cite{runjob05},
    SOKOL \cite{sokol} 
    (iron group),
   as well as
   fluxes from indirect measurements (iron group) by
   KASCADE electrons and muons for two hadronic interaction models
   \cite{ulrichapp} and
   EAS-TOP \cite{eastopspec}.
   The line indicates the spectrum according to the \modell.}
   \label{elementspekfe}
\end{figure}

A significant step forward in the understanding of the origin of cosmic rays
are measurements of energy spectra for individual elements or at least
groups of elements.
Up to about a PeV direct measurements have been performed with instruments
above the atmosphere. As examples, results for primary protons, helium, and
iron nuclei are compiled in Figs.~\ref{elementspekp} to \ref{elementspekfe}.
Recently, also indirect measurements of elemental groups became possible.

A special class of events, the unaccompanied hadrons were investigated by the
EAS-TOP and KASCADE experiments \cite{eastopsh,kascadesh}. Simulations reveal
that these events, where only one hadron is registered in a large calorimeter,
are sensitive to the flux of primary protons. The derived proton fluxes agree
with the results of direct measurements as can be inferred from
\fref{elementspekp}, indicating a reasonably good understanding of the hadronic
interactions in the atmosphere for energies below 1~PeV.

At higher energies a breakthrough has been achieved by the KASCADE experiment.
Measuring simultaneously the electromagnetic and muonic component of air
showers and unfolding the two dimensional shower size distributions, the energy
spectra of five elemental groups have been derived \cite{ulrichapp}.
In order to estimate the influence of the hadronic interaction models used in
the simulations, two models, namely QGSJET~01 and SIBYLL \cite{sibyll21}, have
been applied to interpret the measured data. It turns out that the all-particle
spectra obtained agree satisfactory well within the statistical errors. For both
interpretations the flux of light elements exhibits individual \knees.
The absolute flux values differ by about a factor of two or three between the
different interpretations.
However, it is evident that the \knee in the all-particle spectrum is caused by
a depression of the flux of light elements. The KASCADE results are illustrated
in Figs.~\ref{elementspekp} to \ref{elementspekfe}.

In the figures also results from other air shower experiments are shown.
EAS-TOP derived spectra from the simultaneous observation of the
electromagnetic and muonic components.
HEGRA used an imaging \Cerenkov telescope system to derive the primary proton
flux \cite{hegrap}.
Spectra for protons and helium nuclei are obtained from emulsion chambers
exposed at Mts. Fuji and Kanbala \cite{mtfujip}.
The Tibet group performs measurements with a burst detector as well as with
emulsion chambers and an air shower array \cite{tibetbdp,tibetasg03}.

Over the wide energy range depicted, the measurements seem to follow power laws
with a cut-off at high energies.  The spectra according to the \modell are
indicated in the figures as lines. It can be recognized that the measured
values are compatible with cut-offs at energies proportional to the nuclear
charge $\hat{E}_Z=Z\cdot4.5$~PeV.

\section{Conclusion and outlook}
In the last decade substantial progress has been achieved and the knowledge
about galactic cosmic rays has been significantly increased.
The all-particle energy spectrum is reasonably well known.
For the first time, spectra for groups of elements could be derived from air
shower measurements.
The observed spectra seem to exhibit cut-offs proportional to the nuclear
charge. The increase of the mean logarithmic mass derived from air shower
observations seems to be compatible with subsequential cut-offs for individual
elements.
However, the interpretation of the measurements is still limited by the
uncertainties of the description of hadronic interactions in the atmosphere.
The implications of recent measurements on the contemporary
understanding of the origin of the \knee have been discussed elsewhere
\cite{ecrsreview}.

With the recent measurements of TeV $\gamma$-rays from supernova remnants
\cite{hesssnr} a new window has been opened, and the hypothesis of cosmic-ray
acceleration in supernova remnants is tested directly.

With the KASCADE-Grande experiment \cite{grande}, investigations of the energy
spectra for groups of elements will be extended into the region of the expected
iron \knee ($\sim100$~PeV) and up to the second \knee ($\sim400$~PeV).  In
particular, the inclusion of the hadronic component will help to improve the
interaction models at these energies.

A new technique to measure air showers is about to be established: The LOPES
experiment has registered first air showers observing geosynchrotron emission
in the radio frequency range from 40 to 80~MHz \cite{radionature}.

The presently largest balloon borne detector, the TRACER experiment
\cite{tracer05}, almost reaches the energy region of indirect measurements.
Successor experiments will provide twofold progress for the understanding of
the origin of galactic cosmic rays: an extension of the measurements of the
ratio of secondary to primary nuclei to energies approaching the \knee will
improve the knowledge about the propagation of cosmic rays and the extension of
energy spectra with individual element resolution towards the air shower regime
will provide useful information to improve hadronic interaction models.

\section*{References}
%\bibliographystyle{florenz}
%\bibliography{cr}% Produces the bibliography via BibTeX.

\begin{thebibliography}{100}
\expandafter\ifx\csname urlstyle\endcsname\relax
  \providecommand{\doi}[1]{doi:\discretionary{}{}{}#1}\else
  \providecommand{\doi}{doi:\discretionary{}{}{}\begingroup
  \urlstyle{rm}\Url}\fi

\bibitem{origin}
J.R. {H\"orandel}, \emph{Astropart. Phys.} \textbf{21}, 241 (2004).

\bibitem{berezhko}
E.G. Berezhko \& L.T. Ksenofontov, \emph{JETP} \textbf{89}, 391 (1999).

\bibitem{stanev}
T.~Stanev \emph{et~al.}, \emph{Astron. \& Astroph.} \textbf{274}, 902 (1993).

\bibitem{kobayakawa}
K.~Kobayakawa \emph{et~al.}, \emph{Phys. Rev. D} \textbf{66}, 083004 (2002).

\bibitem{sveshnikova}
L.G. Sveshnikova \emph{et~al.}, \emph{Astron. \& Astroph.} \textbf{409}, 799
  (2003).

\bibitem{wolfendale}
A.D. Erlykin \& A.W. Wolfendale, \emph{J. Phys. G: Nucl. Part. Phys.}
  \textbf{27}, 1005 (2001).

\bibitem{voelk}
H.J. {V\"olk} \& V.N. Zirakashvili, \emph{Proc. 28th Int. Cosmic Ray Conf.,
  Tsukuba} \textbf{4}, 2031 (2003).

\bibitem{bednarek}
W.~Bednarek \& R.J. Protheroe, \emph{Astropart. Phys.} \textbf{16}, 397 (2002).

\bibitem{ptuskin}
S.V. Ptuskin \emph{et~al.}, \emph{Astron. \& Astroph.} \textbf{268}, 726
  (1993).

\bibitem{kalmykov}
N.N. Kalmykov \& A.I. Pavlov, \emph{Proc. 26th Int. Cosmic Ray Conf., Salt Lake
  City} \textbf{4}, 263 (1999).

\bibitem{ogio}
S.~Ogio \& F.~Kakimoto, \emph{Proc. 28th Int. Cosmic Ray Conf., Tsukuba}
  \textbf{1}, 315 (2003).

\bibitem{roulet}
E.~Roulet, \emph{Int. J. Mod. Phys. A} \textbf{19}, 1133 (2004).

\bibitem{swordy}
S.P. Swordy, \emph{Proc. 24th Int. Cosmic Ray Conf., Rome} \textbf{2}, 697
  (1995).

\bibitem{lagutin}
A.A. Lagutin \emph{et~al.}, \emph{Nucl. Phys. B (Proc. Suppl.)} \textbf{97},
  267 (2001).

\bibitem{plaga}
R.~Plaga, \emph{New Astronomy} \textbf{7}, 317 (2002).

\bibitem{wick}
S.D. Wick \emph{et~al.}, \emph{Astropart. Phys.} \textbf{21}, 125 (2004).

\bibitem{dar}
A.~Dar, preprint astro-ph/0408310 (2004).

\bibitem{tkaczyk}
S.~Karakula \& W.~Tkaczyk, \emph{Astropart. Phys.} \textbf{1}, 229 (1993).

\bibitem{dova}
M.T. Dova \emph{et~al.}, preprint astro-ph/0112191 (2001).

\bibitem{wigmans}
R.~Wigmans, \emph{Astropart. Phys.} \textbf{19}, 379 (2003).

\bibitem{candia}
J.~Candia \emph{et~al.}, \emph{Astropart. Phys.} \textbf{17}, 23 (2002).

\bibitem{nikolsky}
S.I. Nikolsky \emph{et~al.}, \emph{Phys. Atomic Nuclei} \textbf{63}, 1799
  (2000).

\bibitem{petrukhin}
A.A. Petrukhin, \emph{Phys. Atom. Ncl.} \textbf{66}, 517 (2003).

\bibitem{kazanas}
D.~Kazanas \& A.~Nikolaidis, \emph{Gen. Rel. Grav.} \textbf{35}, 1117 (2001).

\bibitem{hesssnr}
F.~Aharonian \emph{et~al.}, \emph{Nature} \textbf{432}, 75 (2004).

\bibitem{kascade-points}
T.~Antoni \emph{et~al.}, \emph{Astrophys. J.} \textbf{608}, 865 (2004).

\bibitem{cris-abundance}
M.~Wiedenbeck \emph{et~al.}, \emph{Proc. 28th Int. Cosmic Ray Conf., Tsukuba}
  \textbf{4}, 1899 (2003).

\bibitem{cris-time}
N.E. Yanasak \emph{et~al.}, \emph{Astrophys. J.} \textbf{563}, 768 (2001).

\bibitem{kascade-aniso}
T.~Antoni \emph{et~al.}, \emph{Astrophys. J.} \textbf{604}, 687 (2004).

\bibitem{grigorov}
Grigorov \etal~after T.~Shibata, \emph{Nucl. Phys. B (Proc. Suppl.)}
  \textbf{75A}, 22 (1999).

\bibitem{jaceefe}
K.~Asakimori \emph{et~al.}, \emph{Proc. 24th Int. Cosmic Ray Conf., Rome}
  \textbf{2}, 707 (1995).

\bibitem{runjob05}
V.A. Derbina \emph{et~al.}, \emph{Astrophys. J.} \textbf{628}, L41 (2005).

\bibitem{sokol}
I.P. Ivanenko \emph{et~al.}, \emph{Proc. 23rd Int. Cosmic Ray Conf., Calgary}
  \textbf{2}, 17 (1993).

\bibitem{agasa}
M.~Takeda \emph{et~al.}, \emph{Astropart. Phys.} \textbf{19}, 447 (2003).

\bibitem{akeno1}
M.~Nagano \emph{et~al.}, \emph{J. Phys. G: Nucl. Part. Phys.} \textbf{10}, 1295
  (1984).

\bibitem{akeno20}
M.~Nagano \emph{et~al.}, \emph{J. Phys. G: Nucl. Part. Phys.} \textbf{18}, 423
  (1984).

\bibitem{auger05}
AUGER Collaboration~P. Sommers \emph{et~al.}, astro-ph/0507150 (2005).

\bibitem{basjemas}
S.~Ogio \emph{et~al.}, \emph{Astrophys. J.} \textbf{612}, 268 (2004).

\bibitem{blanca}
J.W. Fowler \emph{et~al.}, \emph{Astropart. Phys.} \textbf{15}, 49 (2001).

\bibitem{casae}
M.A.K. Glasmacher \emph{et~al.}, \emph{Astropart. Phys.} \textbf{10}, 291
  (1999).

\bibitem{dice}
S.P. Swordy \& D.B. Kieda, \emph{Astropart. Phys.} \textbf{13}, 137 (2000).

\bibitem{eastope}
M.~Aglietta \emph{et~al.}, \emph{Astropart. Phys.} \textbf{10}, 1 (1999).

\bibitem{flyseye}
D.J. Bird \emph{et~al.}, \emph{Astrophys. J.} \textbf{424}, 491 (1994).

\bibitem{haverahpark91}
M.A. Lawrence \emph{et~al.}, \emph{J. Phys. G: Nucl. Part. Phys.} \textbf{17},
  733 (1991).

\bibitem{haverahpark03}
M.~Ave \emph{et~al.}, \emph{Astropart. Phys.} \textbf{19}, 47 (2003).

\bibitem{hegraairobic}
F.~Arqueros \emph{et~al.}, \emph{Astron. \& Astroph.} \textbf{359}, 682 (2000).

\bibitem{hiresmia}
T.~Abu-Zyyad \emph{et~al.}, \emph{Astrophys. J.} \textbf{557}, 686 (2001).

\bibitem{hiresi}
R.U. Abbasi \emph{et~al.}, \emph{Phys. Rev. Lett.} \textbf{92}, 151101 (2004).

\bibitem{hiresii}
R.U. Abbasi \emph{et~al.}, \emph{Astropart. Phys.} \textbf{23}, 157 (2005).

\bibitem{ulrichapp}
H.~Ulrich \emph{et~al.}, astro-ph/0505413 (2005).

\bibitem{hknie}
J.R. {H\"orandel} \emph{et~al.}, \emph{Proc. 26th Int. Cosmic Ray Conf., Salt
  Lake City} \textbf{1}, 337 (1999).

\bibitem{rothnn}
T.~Antoni \emph{et~al.}, \emph{Astropart. Phys.} \textbf{16}, 245 (2002).

\bibitem{msu}
Y.A. Fomin \emph{et~al.}, \emph{Proc. 22nd Int. Cosmic Ray Conf., Dublin}
  \textbf{2}, 85 (1991).

\bibitem{mtnorikura}
N.~Ito \emph{et~al.}, \emph{Proc. 25th Int. Cosmic Ray Conf., Durban}
  \textbf{4}, 117 (1997).

\bibitem{sugar}
L.~Anchordoqui \& H.~Goldberg, \emph{Phys. Lett. B} \textbf{583}, 213 (2004).

\bibitem{tibetasg00}
M.~Amenomori \emph{et~al.}, \emph{Phys. Rev. D} \textbf{62}, 072007 (2000).

\bibitem{tibetasg03}
M.~Amenomori \emph{et~al.}, \emph{Proc. 28th Int. Cosmic Ray Conf., Tsukuba}
  \textbf{1}, 143 (2003).

\bibitem{tunka04}
D.~Chernov \emph{et~al.}, astro-ph/0411139 (2004).

\bibitem{yakutsk5001000}
A.V. Glushkov \emph{et~al.}, \emph{Proc. 28th Int. Cosmic Ray Conf., Tsukuba}
  \textbf{1}, 389 (2003).

\bibitem{pg}
J.R. {H\"orandel}, \emph{Astropart. Phys.} \textbf{19}, 193 (2003).

\bibitem{casam}
M.A.K. Glasmacher \emph{et~al.}, \emph{Astropart. Phys.} \textbf{12}, 1 (1999).

\bibitem{chacaltaya}
C.~Aguirre \emph{et~al.}, \emph{Phys. Rev. D} \textbf{62}, 032003 (2000).

\bibitem{eastop-knee}
M.~Aglietta \emph{et~al.}, \emph{Astropart. Phys.} \textbf{21}, 583 (2004).

\bibitem{eastop-macro-lna}
M.~Aglietta \emph{et~al.}, \emph{Astropart. Phys.} \textbf{20}, 641 (2004).

\bibitem{hegracrt}
K.~{Bernl\"ohr} \emph{et~al.}, \emph{Astropart. Phys.} \textbf{8}, 253 (1998).

\bibitem{kascadehm}
J.R. {H\"orandel} \emph{et~al.}, \emph{Proc. 16th European Cosmic Ray
  Symposium, Alcala de Henares} p. 579 (1998).

\bibitem{spaseamandalna}
K.~Rawlins \emph{et~al.}, \emph{Proc. 28th Int. Cosmic Ray Conf., Tsukuba}
  \textbf{1}, 173 (2003).

\bibitem{jaceemasse}
JACEE collaboration~after T.~Shibata, \emph{Nucl. Phys. B (Proc. Suppl.)}
  \textbf{75A}, 22 (1999).

\bibitem{wiebel}
B.~Wiebel-Soth \emph{et~al.}, \emph{Astron. \& Astroph.} \textbf{330}, 389
  (1998).

\bibitem{buettner}
C.~{B\"uttner} \emph{et~al.}, \emph{Proc. 28th Int. Cosmic Ray Conf., Tsukuba}
  \textbf{1}, 33 (2003).

\bibitem{cacti}
S.~Paling \emph{et~al.}, \emph{Proc. 25th Int. Cosmic Ray Conf., Durban}
  \textbf{5}, 253 (1997).

\bibitem{haverahpark00}
A.A. Watson, \emph{Phys. Rep.} \textbf{333 - 334}, 309 (2000).

\bibitem{hires00}
T.~Abu-Zayyad \emph{et~al.}, \emph{Phys. Rev. Lett.} \textbf{84}, 4276 (2000).

\bibitem{mtlianwang}
M.~Cha \emph{et~al.}, \emph{Proc. 27th Int. Cosmic Ray Conf., Hamburg}
  \textbf{1}, 132 (2001).

\bibitem{spase99}
J.E. Dickinson \emph{et~al.}, \emph{Proc. 26th Int. Cosmic Ray Conf., Salt Lake
  City} \textbf{3}, 136 (1999).

\bibitem{yakutsk}
S.~Knurenko \emph{et~al.}, \emph{Proc. 27th Int. Cosmic Ray Conf., Hamburg}
  \textbf{1}, 177 (2001).

\bibitem{wq}
J.R. {H\"orandel}, \emph{J. Phys. G: Nucl. Part. Phys.} \textbf{29}, 2439
  (2002).

\bibitem{corsika}
D.~Heck \emph{et~al.}, Report FZKA 6019, Forschungszentrum Karlsruhe (1998).

\bibitem{qgsjet}
N.N. Kalmykov \emph{et~al.}, \emph{Nucl. Phys. B (Proc. Suppl.)} \textbf{52B},
  17 (1997).

\bibitem{belovisvhecri}
K.~Belov \emph{et~al.}, \emph{Nucl. Phys. B (Proc. Suppl.)} (Proc. 13th
  ISVHECRI) in press (2005).

\bibitem{isvhecri04wq}
J.R. {H\"orandel}, \emph{Nucl. Phys. B (Proc. Suppl.)} (Proc. 13th ISVHECRI)
  in press (2005).

\bibitem{isvhecri02wwtest}
J.R. {H\"orandel}, \emph{Nucl. Phys. B (Proc. Suppl.)} \textbf{122}, 455
  (2003).

\bibitem{amsp}
J.~Alcaraz \emph{et~al.}, \emph{Phys. Lett. B} \textbf{490}, 27 (2000).

\bibitem{atic03}
H.S. Ahn \emph{et~al.}, \emph{Proc. 28th Int. Cosmic Ray Conf., Tsukuba}
  \textbf{4}, 1833 (2003).

\bibitem{bess00}
T.~Sanuki \emph{et~al.}, \emph{Astrophys. J.} \textbf{545}, 1135 (2000).

\bibitem{caprice98}
M.~Boezio \emph{et~al.}, \emph{Astropart. Phys.} \textbf{19}, 583 (2003).

\bibitem{heat01}
M.A.~Du Vernois \emph{et~al.}, \emph{Proc. 27th Int. Cosmic Ray Conf., Hamburg}
  \textbf{5}, 1618 (2001).

\bibitem{ichimura}
M.~Ichimura \emph{et~al.}, \emph{Phys. Rev. D} \textbf{48}, 1949 (1993).

\bibitem{imax00}
M.~Menn \emph{et~al.}, \emph{Astrophys. J.} \textbf{533}, 281 (2000).

\bibitem{jaceephe}
K.~Asakimori \emph{et~al.}, \emph{Astrophys. J.} \textbf{502}, 278 (1998).

\bibitem{mass99}
R.~Bellotti \emph{et~al.}, \emph{Phys. Rev. D} \textbf{60}, 052002 (1999).

\bibitem{papini}
P.~Papini \emph{et~al.}, \emph{Proc. 23rd Int. Cosmic Ray Conf., Calgary}
  \textbf{1}, 579 (1993).

\bibitem{rich2}
E.~Diehl \emph{et~al.}, \emph{Astropart. Phys.} \textbf{18}, 487 (2003).

\bibitem{ryanp}
M.J. Ryan \emph{et~al.}, \emph{Phys. Rev. Lett.} \textbf{28}, 985 (1972).

\bibitem{smith}
L.H. Smith \emph{et~al.}, \emph{Astrophys. J.} \textbf{180}, 987 (1973).

\bibitem{zatsepinp}
V.I. Zatsepin \emph{et~al.}, \emph{Proc. 23rd Int. Cosmic Ray Conf., Calgary}
  \textbf{2}, 14 (1993).

\bibitem{kascadesh}
T.~Antoni \emph{et~al.}, \emph{Astrophys. J.} \textbf{612}, 914 (2004).

\bibitem{eastopspec}
G.~Navarra \emph{et~al.}, \emph{Proc. 28th Int. Cosmic Ray Conf., Tsukuba}
  \textbf{1}, 147 (2003).

\bibitem{eastopsh}
M.~Aglietta \emph{et~al.}, \emph{Astropart. Phys.} \textbf{19}, 329 (2003).

\bibitem{hegrap}
F.~Aharonian \emph{et~al.}, \emph{Phys. Rev. D} \textbf{59}, 092003 (1999).

\bibitem{chacaltayap}
N.~Inoue \emph{et~al.}, \emph{Proc. 25th Int. Cosmic Ray Conf., Durban}
  \textbf{4}, 113 (1997).

\bibitem{mtfujip}
J.~Huang \emph{et~al.}, \emph{Astropart. Phys.} \textbf{18}, 637 (2003).

\bibitem{tibetbdp}
M.~Amenomori \emph{et~al.}, \emph{Phys. Rev. D} \textbf{62}, 112002 (2000).

\bibitem{tibetasgp}
M.~Amenomori \emph{et~al.}, \emph{Proc. 28th Int. Cosmic Ray Conf., Tsukuba}
  \textbf{1}, 107 (2003).

\bibitem{webberhe}
W.R. Webber \emph{et~al.}, \emph{Proc. 20th Int. Cosmic Ray Conf., Moscow}
  \textbf{1}, 325 (1987).

\bibitem{sibyll21}
R.~Engel \emph{et~al.}, \emph{Proc. 26th Int. Cosmic Ray Conf., Salt Lake City}
  \textbf{1}, 415 (1999).

\bibitem{crn}
D~{M\"uller} \emph{et~al.}, \emph{Astrophys. J.} \textbf{374}, 356 (1991).

\bibitem{heao3}
J.J. Engelmann \emph{et~al.}, \emph{Astron. \& Astroph.} \textbf{148}, 12
  (1985).

\bibitem{juliusson}
E.~Juliusson \emph{et~al.}, \emph{Astrophys. J.} \textbf{191}, 331 (1974).

\bibitem{minagawa}
G.~Minagawa \emph{et~al.}, \emph{Astrophys. J.} \textbf{248}, 847 (1981).

\bibitem{tracer05}
D.~{M\"uller} \emph{et~al.}, \emph{Proc. 29th Int. Cosmic Ray Conf., Pune}
  (2005).

\bibitem{hareyama}
M.~Hareyama \emph{et~al.}, \emph{Proc. 26th Int. Cosmic Ray Conf., Salt Lake
  City} \textbf{3}, 105 (1999).

\bibitem{ecrsreview}
J.R. {H\"orandel}, astro-ph/0501251 (2005).

\bibitem{grande}
G.~Navarra \emph{et~al.}, \emph{Nucl. Instr. \& Meth. A} \textbf{518}, 207
  (2004).

\bibitem{radionature}
H.~Falcke \emph{et~al.}, \emph{Nature} \textbf{435}, 313 (2005).

\end{thebibliography}

\end{document}